\documentclass[%
 reprint,
 amsmath,amssymb,
 aps,
]{revtex4}

\usepackage{graphicx}
\usepackage{dcolumn}
\usepackage{bm}

\begin{document}

\preprint{APS/123-QED}

\title{Designing 
 Hyper-Thin Acoustic Metasurfaces with Membrane Resonators}

\author{Yao-Ting Wang\textsuperscript{1} and Richard V. Craster\textsuperscript{1,2}}
    \email[Correspondence email address: ]{r.craster@imperial.ac.uk}
    \affiliation{\textsuperscript{1}Department of Mathematics, Imperial College London, London, SW7 2AZ, United Kingdom}
     \affiliation{\textsuperscript{2}Department of Mechanical Engineering, Imperial College London, London, SW7 2AZ, United Kingdom}
\date{\today} 

\begin{abstract}
We design extremely-thin acoustic metasurfaces, providing a versatile platform for the manipulation of reflected pressure fields, that are constructed from mass loads and stretched membranes fixed to a periodic rigid framework. These metasurfaces demonstrate deeply subwavelength control and can have thicknesses an order of magnitude less than those based around Helmholtz resonators. 
Each sub-unit of the metasurface is resonant at a frequency tuned geometrically, this tunability provides phase control and using a set of finely tuned membrane resonators we create a phase-grating metasurface. This surface is designed to exhibit all-angle negative reflections with the ratio of wavelength, $\lambda$, to thickness, $h$, of $\lambda/h\approx 23.1$, and to create a flat mirror using the phase profile of an elliptic reflecting mirror. A further important acoustic application is to sound diffusers  and we proceed to design a  deeply subwavelength membrane-based meta-diffuser that can be two orders of magnitude thinner than the operating wavelength, i.e. thickness $\approx\lambda/102$. This paves the way for developing advanced acoustic metasurfaces with applicability to functional acoustic devices in sound-related industries.
\end{abstract}


\maketitle

\section{Introduction} \label{sec:intro}
Metasurfaces, as two-dimensional flat structures built from sub-wavelength components, have been instrumental in the development of flat optical devices to manipulate light \cite{Neshev2018,Kamali2018,NanfangYu2014}. These  devices deform the desired wavefront by manipulating the phase on the scale of a wavelength \cite{NanfangYu2011} and, due to their compact size, offer design advantages.   
To achieve the required phase control this artificial flat structure is usually constructed from  micro-resonators that play the role of illuminated pixels and re-build the light wave in accordance with Huygens' principle. The same concepts in wave physics naturally apply to acoustics as surfaces, in this context, are also readily engineered to have sub-wavelength components. Conventionally in acoustics, metasurfaces consisting of periodically graded phase distribution are achieved by mapping surface impedance profiles with slits \cite{JiajunZhao2013a,JiajunZhao2013}, exploiting space-coiling structures \cite{ZixiangLiang2012,YongLi2012a,XuefengZhu2016,YangboXie2013a,YangboXie2013,li2013reflected,wang2016subwavelength}, or by using Helmholtz resonators with various quality factors \cite{ShilongZhai2015,ChanglinDing2015,YongLi2015,long2018reconfigurable,guo2018manipulating}. Since such metasurfaces provide  additional spatially dependent phase profiles, Snell's law needs to be corrected with regards to the law of momentum conservation; this can result in asymmetric, even negative, reflection \cite{BingyiLiu2016a,BingyiLiu2016,BongyiLiu2017,fan2019tunable} or refraction \cite{YongLi2016a,YadongXu2015}, for all-angle incidences. Apart from anomalous wave deflection, to date, several implementations have been realised for a variety of applications such as acoustic lensing devices \cite{YongLi2012a,Wenqi2014,YongLi2014,qi2018ultrathin}, acoustic perfect absorbers \cite{GuancongMa2014,YongLi2016,jimenez16b,huang2018acoustic,huang2019acoustic} and coupling these to concepts such as rainbow trapping \cite{jimenez17a}, sound energy harvesting \cite{qi2016acoustic,qi2017acoustic}, acoustic invisibility carpets \cite{Faure2016,Esfahlani2016}, and  sound-vortex generators \cite{LipingYe2016,JiangXue2016}. Such surfaces can be readily built using  discrete metamaterial bricks \cite{memoli2017bricks} and more recently programmable metasurfaces have been developed \cite{tian2019programmable}.
 
 \indent However, while most of the existing studies of acoustic metasurfaces are constructed of Helmholtz resonators or coiling-up-space structures, they can suffer from energy dissipation although under some circumstances this is rephrased as an advantage \cite{Gerard_jing_2020}. As sound waves oscillate back-and-forth through the neck of the Helmholtz resonators, or along the space-coiling structure, viscous and thermal losses accumulate  \cite{Ward2015,brandao2020}.
  Acoustic metasurfaces built from space-coiling or Helmholtz resonators facilitate many fascinating phenomena, and create thinner devices than conventional designs, but these surfaces require the physical space to form a resonant chamber or to fold the structural channels and hence size limitations remain. To go one stage further we develop 
  a hyper-thin acoustic metasurface using membrane resonators; this metasurface is a two-dimensional version of locally resonant sonic metamaterials consisting of hard metal beads embedded in soft rubber background \cite{liu2000locally}. Membrane-type metasurfaces have been shown to be a feasible alternative to accomplish extraordinary effects such as perfect sound absorbers \cite{GuancongMa2014} or perfect water-to-air transmission \cite{bok2018metasurface}. Our aim here is to implement a graded  array of thin membranes combined with finely tuned mass loads to design extremely thin   metasurfaces with designed phase profiles.

Phase-graded metasurfaces are used to achieve negative reflection and flat focusing mirrors and we use these effects to illustrate the generality of the metasurfaces proposed. A phase-graded acoustic metasurface is composed of a set of periodically arranged resonator super-cells; within every super-cell, a linear phase distribution ranging from 0 to 2$\pi$ is introduced \cite{BingyiLiu2016a,BingyiLiu2016,BongyiLiu2017,fan2019tunable}. From the generalized laws of reflection the wave reflected by the gradient metasurface exhibits a critical angle which defines the domain for anomalous reflection. Concentrating acoustic energy can also be achieved when the phase profiles throughout the acoustic metasurface are set to be the reflected phase of a conic section curve \cite{YongLi2012a,Wenqi2014,YongLi2014,qi2018ultrathin}; by exploiting the metasurface design, specific phase pixelation is generated by the corresponding resonator unit so that the collective wavefront forms a focusing beam above a ultra-thin mirror.

  \begin{figure}[ht!]
  \centering
  \includegraphics[width=0.48\textwidth]{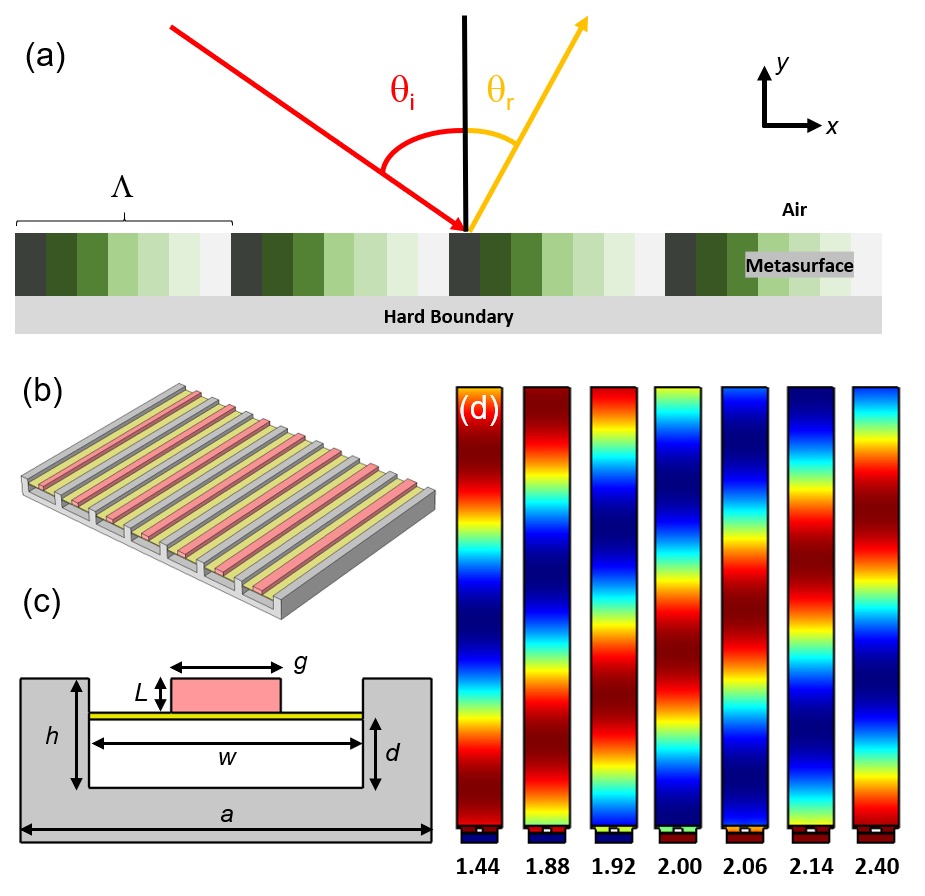}
  \caption{(a) A schematic of a phase-grating acoustic metasurface, where $\theta_i$ and $\theta_r$ denote incident and reflected angles, respectively. (b) An acoustic metasurface consisting of a group of mass loads on stretched membranes; each sub-unit is resonant at different frequency and together they generate the desired phase gradient. (c) The side view of one sub-unit. 
  (d) The scattered acoustic field of the membrane-resonator structures illuminated at a frequency $ 4640$ Hz. The bold numbers give the widths of mass loads, \textit{g}, in millimeters and the corresponding phase profile covers a range of nearly $2\pi$.}
  \label{fig1}
\end{figure}

 As another key application of the metasurfaces proposed here we design diffusers, these are surfaces designed to scatter sound energy equally in all directions to generate a diffusive acoustic space and are commonly used to remove coloration (in optics) and remove echoes in architectural acoustics. These originate from two seminal papers on realising sound diffusion from maximum-length and quadratic-residue sequences proposed by Schroeder \cite{schroeder1975,schroeder1979} after whom these diffusers are named; the critical idea is that simple phase-grating devices can scatter sound into all directions. The conventional configuration to implement phase grating is composed of sound wells with specific depths in terms of specific number sequence \cite{book2009,hargreaves2000surface,d2000diffusor}; each well yields a different resonant frequency and thus the phase of reflected waves can be locally tuned by adjusting geometrical parameters. Therefore, a periodic arrangement of such phase grating panels leads to sound diffusion at the certain frequencies.  Schroeder's work laid the foundation for developing many modern sonic devices in architectural acoustics \cite{cox1994prediction,hargreaves2000surface}, sonography \cite{huang2006producing}, and noise mitigation \cite{mahmoud2014using,wang2015sound,farrehi2016reducing}. 
 Recent studies of metasurfaces have produced several designs based on Helmholtz resonators that have reduced the thickness of diffusers down to several tenths of a wavelength \cite{jimenez2017metadiffusers,zhu2017ultrathin} as realised experimentally \cite{Ballestro2019}.

  \indent This article is structured as follows: firstly we briefly review the concept of generalised Snell's law in a surface with graded phase pattern, then the design of membrane-type resonators and a prototype of phase-grating metasurface are demonstrated. Based on the designed phase-grating metasurface, we theoretically verify its capabilities to generate anomalous reflections and a sonic flat focusing mirror. In Sec. \ref{sec:sound_diff}, a hyper-thin sound diffuser with membrane resonators is evaluated and  compared with conventional diffusers; our design shrinks the thickness down to less than a hundredth of wavelength. Diffusers operating at different frequencies and the effect of material loss from membrane are also discussed. We draw together concluding remarks in Sec. \ref{sec:conclud}.

\section{abnormal reflection and planar focusing} \label{sec:agm}
Following \cite{NanfangYu2011}, we use a generalized law of reflection resulting from graded phase modulation;  Fig.\ref{fig1}a present a schematic of an acoustic gradient metasurface where each super-cell, whose period is $\Lambda$, consists of \textit{N} sub-units with width $\Lambda/N$. If the flat structure satisfies long-wavelength conditions $\lambda/\Lambda \gg 1$, the graded phase is then homogenised and is approximated as a continuous function with the expression $\phi(x)=s2\pi x/\Lambda$, where \textit{s} determines the incident wave directions. This additional phase modulation modifies the law of reflection due to the wave-vector continuity across the reflected boundary. Thus, with obliquely incident plane waves, the generalised law of reflection is given by
\begin{equation}\label{eq1}
     k_0(\sin\theta_r-\sin\theta_i) = \eta,
\end{equation}
\noindent where \(\theta_r\) and \(\theta_i\) are, respectively, the reflected and incident angles. Here  \(\eta\) is given by \(s2\pi/\Lambda\), denoting the gradient of phase modulation given by \(\eta=\nabla \phi\).

  \begin{figure}[ht!]
    \centering
    \includegraphics[width=0.48\textwidth]{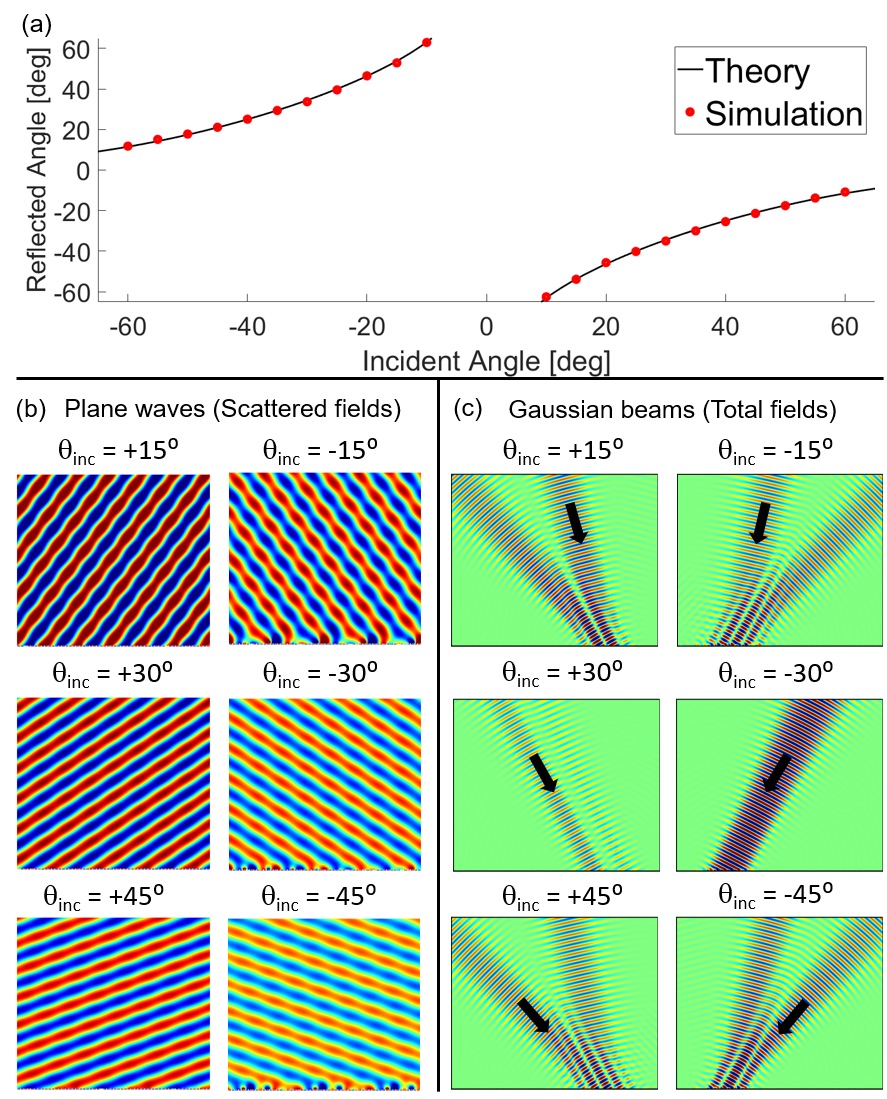}
    \caption{(a) The relation between incident, $\theta_i$, and reflected, $\theta_r$, angle. The black solid lines and red dots account for the values analysed from Eq.\eqref{eq1} and numerical results, respectively for operating frequency 4.64 kHz. (b) Reflected plane waves for incident angles from $\pm15^{\circ}$, $\pm30^{\circ}$ to $\pm45^{\circ}$. The field strengths for negative incidence is weaker as some energy goes into surface waves that are excited. (c) Total field patterns of incident Gaussian beams. The black arrows highlight the directions of incident Gaussian. These figures provide a clearer version of negative reflections. Note that the destructive (constructive) interference occurs at incident angle equal to $30^{\circ}$ (-$30^{\circ}$).}
    \label{fig2}
 \end{figure}

\indent 
 Fig.\ref{fig1}b demonstrates the design of acoustic metasurface composed of membrane resonators. The mass bars, silicon rubber membranes, and rigid framework, are coloured in red, yellow, and grey, respectively. All the cavity chambers are filled with air ($\rho$ = 1.293 kg/m\textsuperscript{3}, c = 343 m/s) and the mass loads are made of epoxy resin ($\rho$ = 1450 kg/m\textsuperscript{3}, $E$ = 3.5 GPa, $\nu$ = 0.33). Both ends of the membrane are fixed on a rigid substrate, all the membranes $\rho$ = 990 kg/m\textsuperscript{3}, $E$ = 5 MPa, $\nu$ = 0.49) are uniformly stretched by a tension equal to 0.4 MPa. As membranes are soft and elastic owing to their high Poisson's ratio,  they can be effectively regarded as springs, that is, the structure forms a series of harmonic resonators that can be characterised simply by spring-mass models \cite{bok2018metasurface}; in Fig.\ref{fig1}(c), we illustrate the side view of a sub-unit cell to label the geometrical parameters which are the lattice constant, $a$, is 10 mm, the cavity depth, $d$, is 2 mm, the cavity width, $w$ is 8 mm, the height of every mass load \textit{L} is 1 mm, and the thickness of membrane is 0.2 mm.
  To tune the phase of reflected waves, the length of mass load is changed as the width \textit{g} varies.
\begin{figure}[htb!]
  \centering
\includegraphics[width=7.5cm]{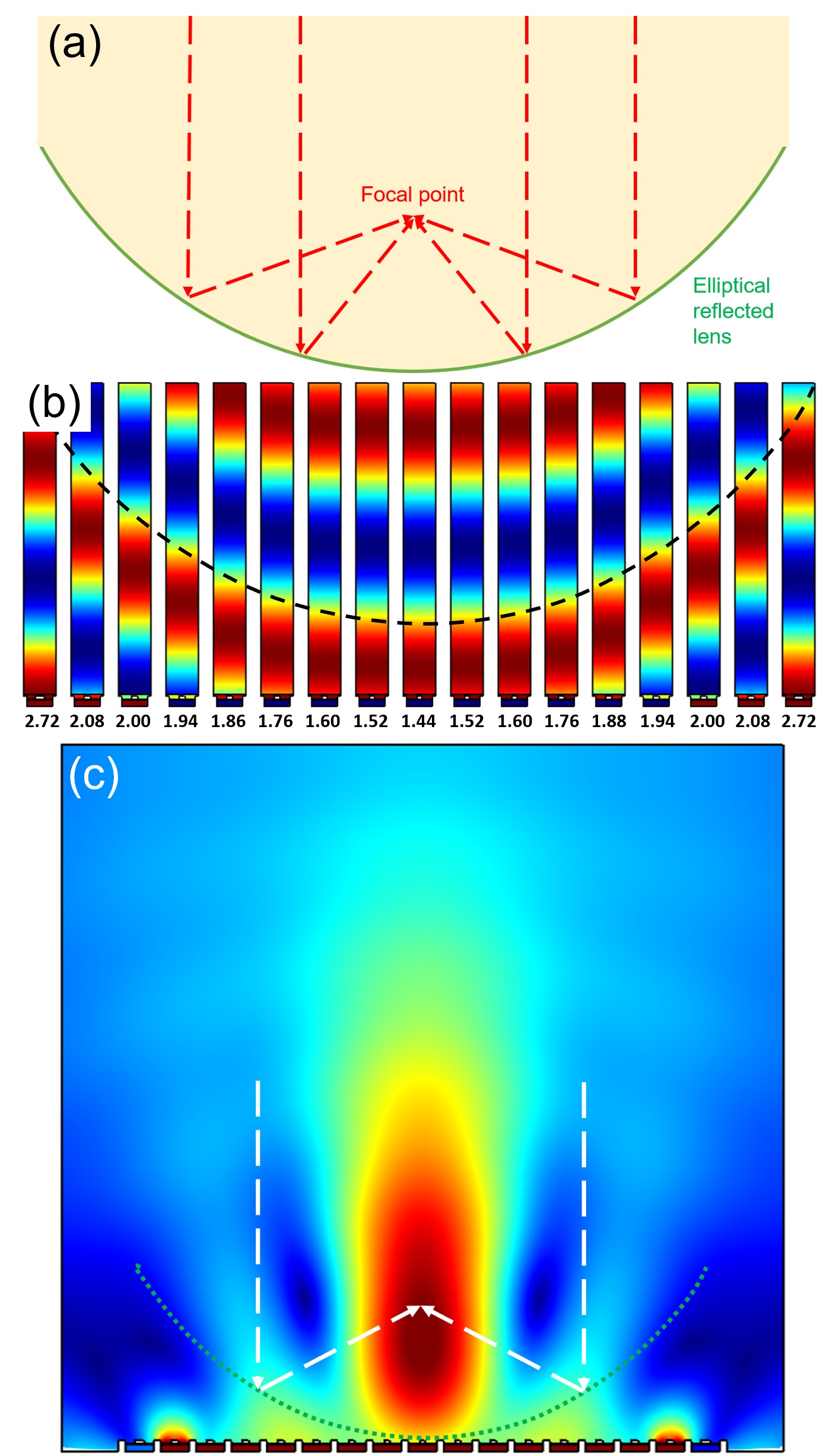}   \caption{(a) Conventional elliptical focusing mirror. (b) The phase profiles that rebuild the phase distribution of a elliptical focusing mirror as constructed by different mass loads whose widths (in millimetres) are written in bold numbers. (c) The absolute value of the scattered pressure field. The white dashed arrows give the trajectory of sound beams.} 
  \label{fig3}
\end{figure}
\begin{figure*}[ht!]
  \centering
  \includegraphics[width=17cm]{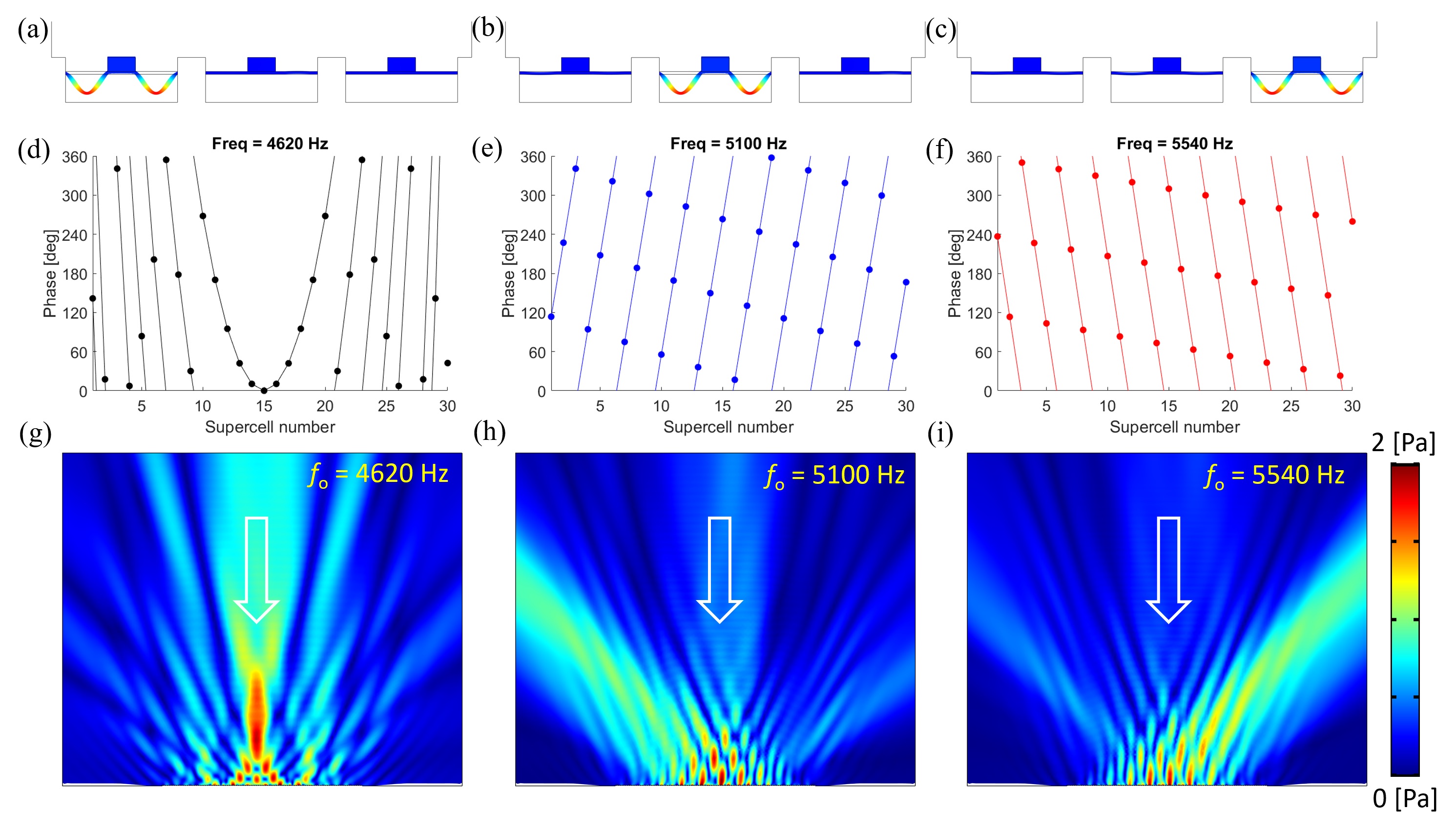}
  \caption{Multi-functional metasurface: The displacement patterns in a period of integrated membrane resonators at chosen frequencies equal to (a) 4.62 kHz, (b) 5.10 kHz, and (c) 5.54 kHz. For each frequency, we design a particular phase profile defined by (d) $k_{0}\times1.2[m]\sqrt{1-(x/0.5[m])^2}$ for planar focusing mirror, (e) $k_{0}\times\sin{45^{\circ}}x$ for positive anomalous reflection, and (f) $-k_{0}\times\sin{45^{\circ}}x$ for negative anomalous reflection, where $k_{0}$ denotes the corresponding wave-vector in air. The absolute field profiles demonstrate three functionalities: (g) planar focusing property at 4.62 kHz, (h) positive anomalous reflection at 5.10 kHz, and (i) negative anomalous reflection at 5.54 kHz. The white arrows depict the direction of incident Gaussian beam.}
  \label{fig4}
\end{figure*}
   A typical set of reflected phases for various \textit{g} values are shown in Fig. \ref{fig1}(d), for operating frequency $f_o = 4640$ Hz, demonstrating the graded phase modulation for such a system; these results come from numerical simulations using COMSOL (a commercial finite element package) \cite{comsol}. The ratio of wavelength to thickness $\lambda/d \approx 23.1$, which validates the hyper-thin feature of membrane-type metasurfaces compared to other designs.\\
\indent Eq.\eqref{eq1} allows us to relate the reflected to incident angles via the expression $\theta_r=\arcsin(\sin\theta_i+\eta/k_0)$ and we cross-verify this relation with finite element simulations in Fig.\ref{fig2}(a). As the range of negative reflections covers virtually every incident angle our design can facilitate nearly omni-directional negative reflections for this hyper-thin acoustic metasurface and this is illustrated in  Fig.\ref{fig2}(b) showing scattered waves, excited by the waves with different incident angles, following the predictions of  \eqref{eq1}. As surface modes can arise, the scattered field strength of incident plane waves, with negative angle, is relatively weak as compared with positive ones. Fig.\ref{fig2}(c) depicts a series of simulations for total pressure fields (both background and scattered fields); this further validates negative reflections for a series of incident angles.
In addition to the abnormal reflections described above, the phase modulation scheme also enables us to create flat focusing mirrors via manipulating the scattered wavefront. There are several designs that can achieve wave focusing; in Fig.\ref{fig3}(a), we exploit an elliptical phase profile to generate planar focusing metasurfaces. By using the same curve, in Fig.\ref{fig3}(b) we then design an approximate phase within 17 units to re-construct the desired wavefront.  Fig.\ref{fig3}(c) illustrates the absolute value of scattered pressure field for an incident 
 acoustic Gaussian beam incident on the metasurface, with the focusing clearly apparent. The green dash line represents the corresponding elliptical phase distribution showing that the position of focal spot follows the prediction in Fig.\ref{fig3}(a).
In Fig.\ref{fig4} we demonstrate a frequency-dependent multi-functional acoustic metasurface that combines both anomalous reflections and planar focusing functionalities \cite{zhu2019multifunctional}. A supercell of multiple resonators with three different excited frequencies are shown in Fig.\ref{fig4}(a)-(c); all structural and material parameters remain the same as earlier.  Fig.\ref{fig4}(d)-(f) show three phase profiles for planar focusing at 4.62 kHz, positive anomalous reflection at 5.10 kHz, and negative anomalous reflection at 5.54 kHz. The numerical results in Fig.\ref{fig4}(g)-(i) validate that multi-functional properties can be easily engineered. Other functionalities, such as energy absorption, can be also integrated into this metasurface, which may add further value to practical applications.

\section{hyper-thin Meta-diffuser} \label{sec:sound_diff}
\begin{figure}[hbt!]
  \centering
  \includegraphics[width=0.48\textwidth]{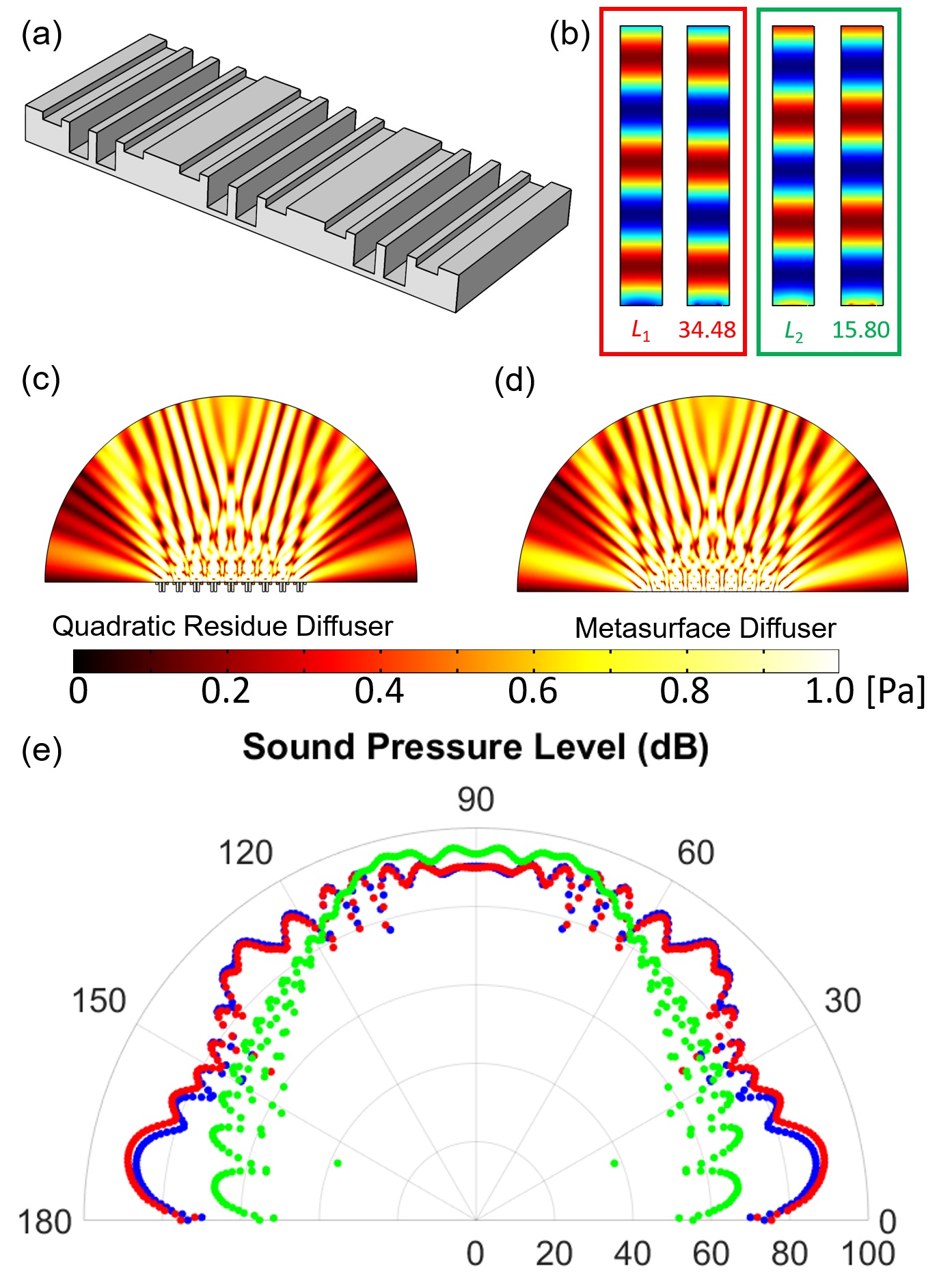}
  \caption{(a) A QRD diffuser, shown for N = 5, where the phases of reflection are defined by the depth of each sound well. (b) The reflected phases, at 550~Hz, from both the sub-units of QRD and membrane resonators with $L_1$ and $L_2$ equal 62.34 and 249.45 mm respectively, which corresponds to mass width of 34.88 and 15.80 mm; the scattered field patterns are illustrated in (c) for the QRD and (d) for the meta-diffuser at 1530~Hz. Both diffusers contain 9 periods and their overall lengths are 4.05 m. The scattered fields are excited by a plane wave normally incident from above. (e) Sound pressure level, at 1530~Hz, for a planar reflector (green), QRD (blue), and meta-diffuser (red), this validates the performance of the QRD and the meta-diffuser}
  \label{fig5}
\end{figure}

In this section, we apply membrane-type metasurfaces to the design of sound diffusers; several designs of ultra-thin meta-diffusers via implementing Helmholtz resonators, or slit-loaded resonator systems, have been proposed \cite{jimenez2017metadiffusers,zhu2017ultrathin,Ballestro2019}, here we further scale the thickness down to $\lambda/100$.  To realise various reflected phases in a quadratic residue diffuser (QRD), each sub-unit of the grating, which is shown in Fig.\ref{fig5}a, is comprised of a set of sound wells with specific depth defined by \cite{schroeder1979}.
    \begin{equation} \label{eq2}
     L_n = (n^2\bmod N )\lambda_d/2N,
    \end{equation}
\noindent where \textit{n} and \textit{N} are the label and the total number of sub-units within a period. Using \textit{N} = 5 as an example, the ratio of $\lambda_d$ to the length of the deepest sub-unit is equal to 2.5, which limits its physical size especially for operation in the low-frequency region; we overcome this limitation by employing membrane resonators to engineer a series of graded phases, originating from QRD, in a hyper-thin metasurface.
\noindent To focus on a typical example, for a design frequency $f_d = 550$ Hz the corresponding depths of wells for a QRD (\textit{N} = 5) are 62.34 and 249.45 mm. The design frequency, which is an optimised value for a better diffuser performance, is not necessarily equal to the operating frequency \cite{cox2003schroeder} and we will discuss broadband versus narrowband performance later. 
 Based on the reflected phases given by the QRD, we numerically calculate the phase from membrane resonators using mass widths $g_1$ and $g_2$ equal to 34.48 and 15.80 mm, respectively; in Fig.\ref{fig4}b, we see that the reflected phase distributions from the conventional QRD and the metasurface are in close agreement with each other. The geometry of a sub-unit is similar to that shown  in Fig.\ref{fig1}c , but with structural parameters: lattice constant \textit{a=} 90 mm, cavity depth \textit{d=} 1 mm, the cavity width \textit{w=} 63 mm and the height of every mass load \textit{L} and the membrane  thickness remain unchanged. All the material parameters are consistent with those presented earlier, except that the tension applied on membranes is increased to 0.2 MPa.
\noindent Fig. \ref{fig5}c-d demonstrate the field profiles of conventional Schroeder diffusers and meta-diffusers at 1.53 kHz. Despite slight discrepancies between two figures, both cases spread sound waves into all directions; the ratio of incident wavelength to thickness, $\lambda_o/h$, for the meta-diffuser significantly increases to approximately 102, which is, to the best of our knowledge, by far the largest magnitude for a diffuser. To evaluate efficiency, a polar plot of scattered sound pressure level ($L_p=20\log_{10}(p_s/p_b)$ with reference pressure in air $p_b = 20 \mu$Pa) is shown in Fig.\ref{fig5}e for a hard wall, Schroeder diffuser, and the corresponding meta-diffuser. In contrast to the directional reflected field from the hard wall, both diffusers exhibit directionality closer to a semi-circle which would provide perfect all-angle diffusion.\\
\indent We further quantitatively characterise the performance of the meta-diffuser via a normalized diffusion coefficient defined by \cite{jimenez2017metadiffusers}
\begin{equation}\label{eq3}
    \delta_{nor}=\frac{\delta-\delta_{ref}}{1-\delta_{ref}}.
\end{equation}
\noindent Here $\delta$ and $\delta_{ref}$ refer to the diffusion coefficients of the diffuser and the hard wall respectively; in particular $\delta$ is numerically acquired by the following equation:
\begin{equation}\label{eq4}
    \delta=\frac{[\begin{matrix} \int_{0}^{\pi} I(\theta)d{\theta} \end{matrix}]^2
                 -\begin{matrix} \int_{0}^{\pi} [I(\theta)]^2d{\theta} \end{matrix}}
                 {\begin{matrix} \int_{0}^{\pi} [I(\theta)]^2d{\theta} \end{matrix}}.
\end{equation}
\begin{figure}[hbt!]
  \centering
  \includegraphics[width=0.48\textwidth]{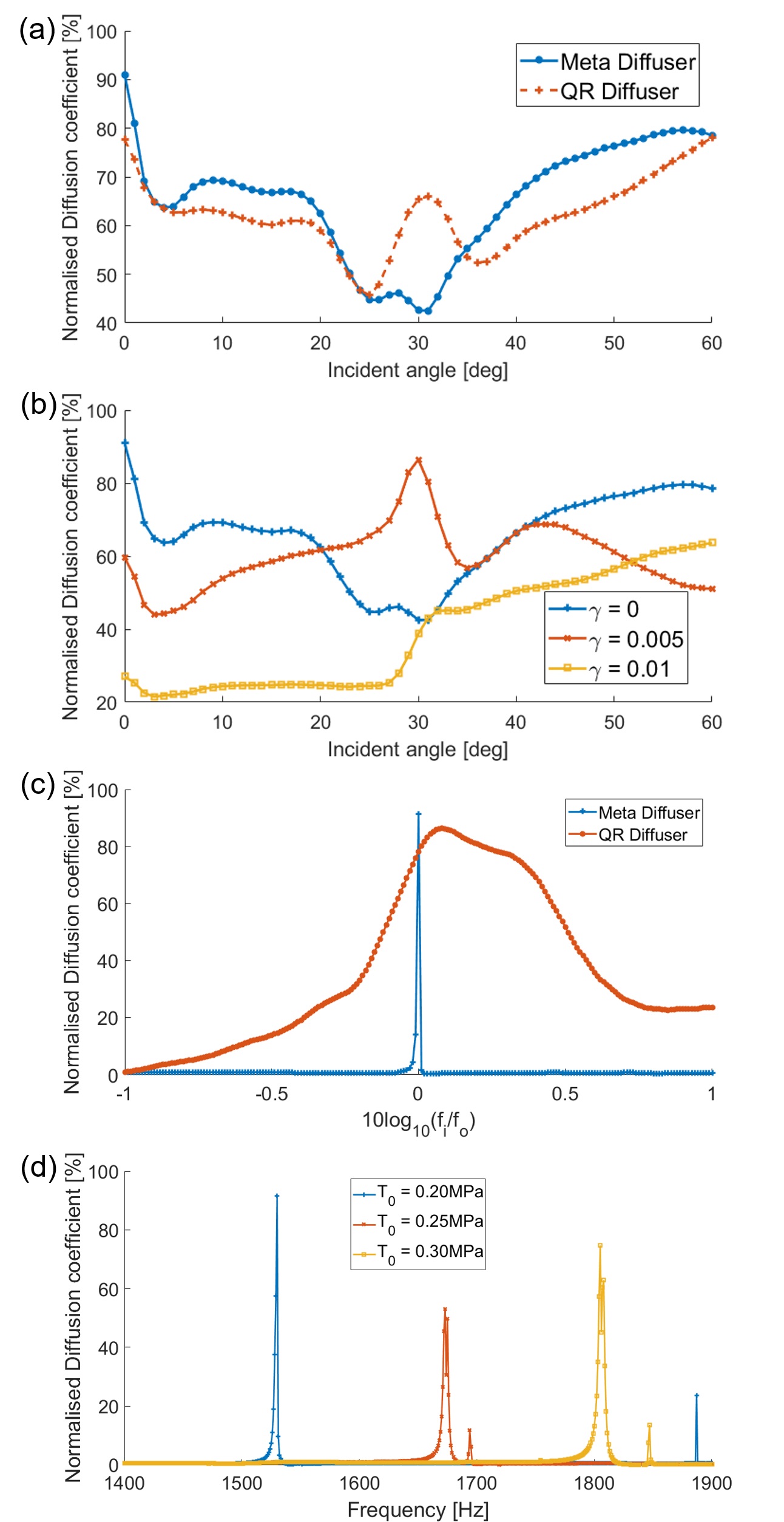}
  \caption{(a) A comparison between QRD and the meta-diffuser. Overall the meta-diffuser performance is better than conventional QRD except for a small range of incident angles between 25 and 35 degrees. (b) The diffusion coefficients showing the effect of material loss. (c) The dispersion curves of diffusion coefficients reflect the resonant feature at the operating frequency 1.53 kHz. (d) Realising reconfigurable meta-diffusers via the tension tunablity of electrically actuated elastomers. For various in-plane tensions, the figure shows that the frequency responses of meta-diffusers can be freely tailored. }
  \label{fig6}
\end{figure}
\noindent where $I(\theta)$ is the intensity function of angles. If the wave reflection is the same as that from a hard wall then this coefficient becomes zero indicating no diffusion effect; in contrast, if the reflected wave is uniformly dispersed over the entire space, the diffusion coefficient equals one \cite{jimenez2017metadiffusers,cox2006tutorial,zhu2017ultrathin}. Considering the incident angle from 0 to 60 degrees, Fig.\ref{fig6}(a) shows $\delta_{nor}$ for the QRD and meta-diffuser. For small-angles of incidence the diffusion coefficient of the meta-diffuser improves over that of the QRD and maintains comparable performance until the angle reaches approximately 25 degrees, and then again becomes greater after 35 degrees. Fig.\ref{fig6}(b) investigates the effect caused from the loss in membranes. As expected, a highly lossy membrane will greatly affect the magnitude of $\delta_{nor}$ but it becomes comparable to the conventional QRD when the loss factor is sufficiently weak: practically, this condition can be achieved by employing commercial low-loss membranes. The diffusion coefficients are strongly affected by the magnitude of loss because it operates around a sharp resonant peak. In other words, when the loss factor rises, the response frequency deviates and the energy dissipation increases, which leads to a noticeable change of the coefficients. However, the design for higher operating frequency may ease the situation despite the decrease of $\lambda_o/h$ ratio (See Appendix B for more discussions). Fig.\ref{fig6}(c) demonstrates the diffusion coefficients of normal incidence versus incident frequencies. It further proves our argument regarding the performance as a sharp peak at $f_o$ is seen. In addition, in Fig.\ref{fig6}(d) we propose a reconfigurable meta-diffuser by introducing a tunable in-plane tension in electrically actuated elastomers \cite{pelrine2000high}, by using different in-plane tensions, the diffuser responses vary from 1.53 kHz to 1.81 kHz, indicating reconfigurable properties.

\section{conclusion} 
\label{sec:conclud}
We have presented simulations for hyper-thin acoustic metasurfaces constructed from membrane resonators. In comparison with acoustic metasurfaces based upon space coiling, or Helmholtz resonators, these membrane resonator metasurfaces have further reduced the structural size, and yet still maintain efficiency; due to recent progress in 3D printing such designs are within practical reach and we anticipate this work will encourage experimentalists to build and test these designs. There is substantial generality in the approach taken here and we demonstrate this, 
for  a thickness $\approx \lambda /23$,  by creating phenomena such as all-angle reflections and flat focusing; this may inspire devices relevant to portable sound devices. A hyper-thin meta-diffuser, that can spread incoming sound into nearly all directions, is designed with an even more extreme slenderness of $\approx \lambda /102$, and its performance even with material loss, appears to rival that of conventional diffusers; 
 this could impact upon echo mitigation and noise control in architectural acoustics.

 


\begin{acknowledgments}
Y.T.W thanks Prof. Guancong Ma and Dr. Min Yang for their valuable suggestions. R.V.C and Y.T.W are funded by the UK Engineering and Physical Sciences Research Council (EP/T002654/1). R.V.C. also acknowledges funding from the ERC H2020 FETOpen project BOHEME.
\end{acknowledgments}

\appendix

\section{Setting of Numerical Simulations}\label{sec:app1}
Numerical simulations for anomalous reflections are performed using the ``Pressure Acoustics" (PA) and ``Solid Mechanics" (SM) modules, with both modules connected by the ``Acoustic-Structure Boundary", in the Multiphysics option for COMSOL Multiphysics v5.5, a commercial  finite-element solver. In frequency-domain calculations both ends of the membranes are fixed to the rigid frame via ``Fixed Constraint" in the SM module. The tension and damping are added by implementing ``Initial Stress and Strain" and ``Damping" respectively in 
 the ``Linear Elastic Material" section. The evaluation domain is surrounded by a hard boundary on the bottom, a perfectly matched layer on the top, and periodic boundary conditions on both sides. The  ``background pressure field" is exploited to introduce a plane wave source. Finally, geometric non-linearity is included 
  before executing the script.

\section{More on Meta-Diffusers}\label{sec:app2}
\begin{figure}[t]
  \centering
  \includegraphics[width=0.48\textwidth]{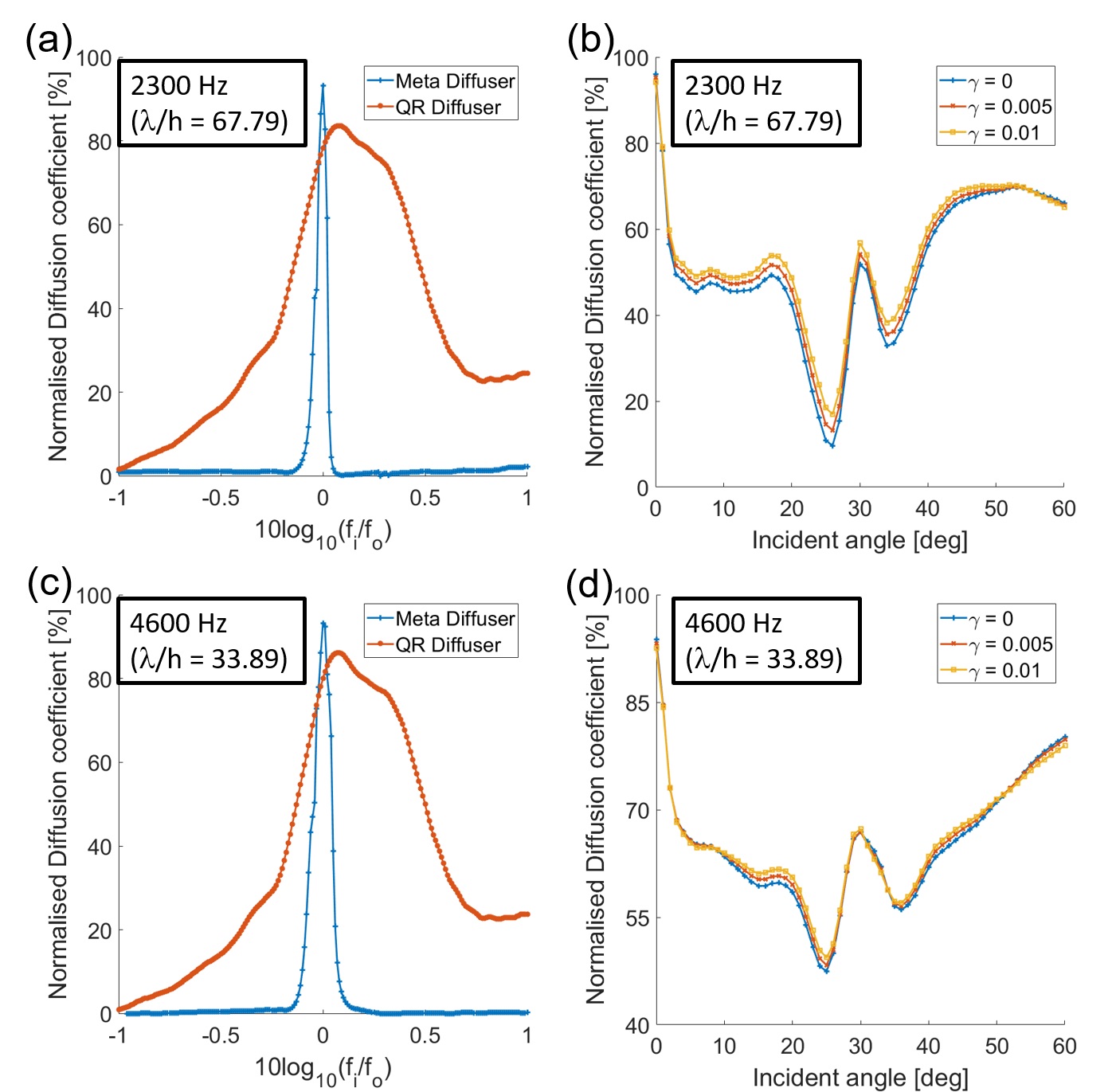}
  \caption{The diffusion coefficients versus (a) frequencies and (b) incident angles at $f_o$ = 2.3 kHz. In comparison with the main text, the former demonstrates a similar but wider resonant peak. This feature results in a insensitive diffusion coefficients as shown in (b). Similar argument applies to the case $f_o$ = 4.6 kHz in (c) and (d).}
  \label{fig7}
\end{figure}

\begin{table}[b]
\begin{ruledtabular}
\begin{tabular}{cccccc}
\centering
\textrm{$f_o$[kHz]}&
\textrm{$f_d$[kHz]}&
\textrm{a[mm]}&
\textrm{$g_1$[mm]}&
\textrm{$g_2$[mm]}&
\textrm{$T_0$[MPa]}\\
\colrule
2.3 & 0.82 & 60 & 19.82 & 20.00 & 0.3\\
4.6 & 1.65 & 30 & 7.28 & 7.46 & 0.3\\
\end{tabular}
\end{ruledtabular}
\caption{\label{tab:tab1}
The parameters for the diffusers operating at different frequency.}
\end{table}

The membrane-type meta-diffuser can be modified to opertate at other frequencies by altering 
 the structural and material settings. Table~\ref{tab:tab1} lists the parameters for the diffusers operating at \textit{$f_o$} = 2.3 kHz ($\lambda_o/h \approx 67.79$) and 4.6 kHz ($\lambda_o/h \approx 33.89$); parameters not listed in Tab.\ref{tab:tab1} remain unaltered from the main text. Fig.\ref{fig7} shows that both cases are fairly robust as loss factors vary as the bandwidth of both resonances are wider ($\Delta f_{4600}$ = 91.15 Hz, $\Delta f_{2300}$ = 26.99 Hz) whereas that shown in the main text has $\Delta f_{1530}$ = 3.51 Hz. In other words, there is a trade-off between bandwidth and the  $\lambda_o/h$ value. It is worth noting that, even though the $\lambda_o/h$ ratio decreases in higher operating frequency, the membrane-type acoustic metasurfaces are still thinner than most other proposed structures. Additionally, compared to the meta-diffusers  \cite{jimenez2017metadiffusers,zhu2017ultrathin}, this design exhibits a nearly perfect performance despite having a relatively narrow bandwidth. In brief, the membrane-type metasurface proposed here may provide another platform of designing novel metasurface-based applications, or a feasible approach of improving current acoustic devices.

\nocite{*}

\providecommand{\noopsort}[1]{}\providecommand{\singleletter}[1]{#1}%

\end{document}